\iftrue
\documentclass[preprint]{aastex}
\slugcomment{Accepted by the Astronomical Journal on 03/31/2013}
\else
\documentclass[iop]{emulateapj}
\usepackage{apjfonts, epsfig}

\include{ms_draft_macros} 
\clearpage
\fi

\usepackage{amsmath}
\usepackage{graphicx}
\usepackage{color}
\usepackage{soul}
\definecolor{myyellow}{rgb}{1,1,0}
\definecolor{myskyblue}{rgb}{0.8,1,1}

\newcommand{\s}[1]{{\sigma^2_{#1}}}

\shorttitle{Period Error Estimation for the Kepler Eclipsing Binary Catalog}
\shortauthors{Mighell}

\begin{document}

\title{Period Error Estimation for the Kepler Eclipsing Binary Catalog}

\author{Kenneth J. Mighell}
\affil{National Optical Astronomy Observatory, 950 North Cherry Avenue, Tucson, AZ 85719}

\and

\author{Peter Plavchan}
\affil{NASA Exoplanet Science Institute, California Institute of Technology,
Pasadena, CA 91125}

\begin{abstract}
The 
Kepler Eclipsing Binary Catalog (KEBC)
describes 2165 eclipsing binaries identified in the 115 deg$^2$ Kepler Field based
on observations from Kepler quarters Q0, Q1, and Q2.
The periods in the KEBC are given in units of days out to six decimal places but no
period errors are provided.
We present the PEC ({\em{P\,}}eriod {\em{E\,}}rror {\em{C\,}}alculator) algorithm which can
be used to estimate the period errors of strictly periodic variables observed by the Kepler Mission.
The PEC algorithm is based on propagation of error theory and assumes that observation of 
every light curve peak/minimum in a long time-series observation can be unambiguously identified.
The  PEC 
algorithm can be efficiently programmed using just a few lines of 
C computer language code.
The PEC algorithm was used to develop a simple model 
which provides period error estimates for eclipsing binaries in the KEBC
with periods less than 62.5 days:
$\log \sigma_P \approx {-5.8908 + 1.4425\!\left(1 +  \log P\right)}$,
where $P$ is the period of an eclipsing binary in the KEBC in units of days.
KEBC systems with periods $\geq$62.5 days have KEBC period errors of $\sim$0.0144 days.
Periods and period errors of 7 eclipsing binary systems in the KEBC
were measured using 
the NASA Exoplanet Archive Periodogram Service and compared to 
period errors estimated using the PEC algorithm.
\end{abstract}


\newpage
\section{INTRODUCTION}

Pr{\v s}a et al.\ (\citeyear{Prsa+2011})
produced the Kepler Eclipsing Binary Catalog (KEBC)
which contained 1879 unique eclipsing and ellipsoidal binary systems
identified in the 115 deg$^2$ Kepler Field using the 
first Kepler data release \citep[and references therein]{Borucki+2011}
which covered the first 44 days of the operation of the Kepler Mission (Kepler quarters Q0 and Q1).
For each object (binary system), the KEBC provides 
the {{Kepler}} ID (KID) number, the period, $P_0$, the ephemeris zero point, BJD$_0$, 
morphology type, and various physical parameters
from the {{Kepler}} Input Catalog. 
For detached and semi-detached
eclipsing binary systems, the KEBC provides the several principal parameters including
the temperature ratio of the two stars,
$T_2/T_1$,
the sum of the fractional radii,
$\rho_1 + \rho_2$,
the radial and tangential components of the eccentricity,
$e \sin(\omega)$ and $e \cos(  \omega)$,
and the sine of the inclination, $\sin(i)$.
For overcontact systems, the KEBC provides 
$T_2/T_1$,
the photometric mass ratio
$q$,
the fillout factor,
and $\sin( i)$.

Slawson et al.\ (\citeyear{Slawson+2011}) updated the Kepler Eclipsing Binary Catalog 
by increasing the baseline nearly fourfold to 125 days
by including the second Kepler data release (Kepler quarter Q2).  Three hundred and eighty-six new systems
were added, and the ephemerides (BJD$_0$ and $P_0$) and principle parameters were recomputed. 
This version of the KEBC contained 2165 objects which is 1.4\% of all Kepler target stars.

The online version of the Kepler Eclipsing Binary Catalog\footnote{{The Kepler Eclipsing Binary Catalog website is\newline
{http://keplerebs.villanova.edu/}}~.
}
is hosted at Villinova University.
The current online version of the KEBC (as of 2013 January 22: Revision 1.96, 2011-06-15)
has information on 2176 systems.  The online version includes figures for each object showing
the raw {{Kepler}} light curve, the detrended light curve, and the phased light curve
with a polynomial model ``fit'' to the data, as determined by the same
neural network analysis which produced the principal parameters.

The statistical properties of the Kepler Eclipsing Binary Catalog were determined using the
EBAI method ({\em{E\,}}clipsing {\em{B\,}}inaries via {\em{A}}rtificial {\em{I\,}}ntelligence; \citealt{Prsa+2008}).  The EBAI
method uses several trained neural networks to determine the principal parameters for every
binary system in the KEBC.
The performance of EBAI on detached eclipsing binaries is described in \S4.2.2. of 
\cite{Prsa+2011}; Figure 10 shows the performance of EBAI method on a test set of 10,000 detached
eclipsing binaries for 5 parameters:
$\sin(i)$,
$e \sin(\omega)$,
$e \cos(\omega)$, 
$\rho_1 + \rho_2$,
and $T_1/T_2$. It was determined that
90\% of all systems had errors smaller than 10\% in all 5 parameters.
Unfortunately, no error analysis was provided for the periods determined using the
test set of 10,000 detached eclipsing binaries.

The periods in the Kepler Eclipsing Binary Catalog are given in units of days out to six decimal places.
One microday is 86.4 milliseconds (ms).
As shown below, the periods for many of KEBC binary systems must have measurement errors that are
significantly greater than 86.4 ms --- especially for the systems that had only a few
complete cycles (periods) in 125 days.
Although no error analysis has been provided for the periods given in the KEBC, we show below
that it is possible to determine reasonable error estimates for periods given in the KEBC
{\em{ex post facto}} without the need to reanalyze the original Kepler light curves.

This article describes how the period errors of strictly periodic variables observed by the Kepler Mission
can be estimated using the new PEC ({\em{P\,}}eriod {\em{E\,}}rror {\em{C\,}}alculator) algorithm which is 
based on propagation of error theory.
The PEC algorithm is described in the next section.
The PEC algorithm is  used in \S 3 to estimate the period errors for periods ranging from
0.1 to 125 days of the 2176 binary systems in the Kepler Eclipsing Binary Catalog.
These theoretical predictions are then compared 
with measured period errors from Kepler observations of 
five detached 
and two semi-detached 
eclipsing binary systems in the KEBC.
The article concludes in \S 4 with a brief discussion of the limitations of using
the PEC algorithm with observations of strictly periodic variables with morphologically complicated
light curves like RR Lyraes.

\section{PERIOD ERROR ESTIMATION} 

In this section we investigate period errors as determined with propagation of error theory
of 10-day uninterrupted time series observations of strictly periodic variables with periods of 7.3 days, 4.7 days, and
2.4 days.  These results are summarized in the last subsection which describes the PEC 
algorithm that can be used to estimate the period errors in the Kepler Eclipsing Binary Catalog.

\subsection{One Cycle}

Let us start with a simple example. Consider a 10-day time series 
of uninterrupted space-based 30-min observations
of a periodic variable with a period $P\!=\!7.3$ days 
and a timing uncertainty of $\sigma\!=\!0.0104$ days (one-half bin width: 15 min).
For the sake of simplicity,
assume that the first peak flux of the light curve occurs during the first observation.
Only a single measurement of a single period is possible.
There will be two peak flux values at $t_1$ and $t_2$.
The estimated period is
\begin{equation}
P_1 = t_2 - t_1\,,
\end{equation}
and the variance of $P_1$ is
\begin{equation}
\sigma^2_{P_1} = \sigma_{t_2}^2 + \sigma_{t_1}^2=2\,\sigma^2\,,
\end{equation}
using propagation of error theory \citep[e.g.,][]{Bevington1969}.
The uncertainty of the period is 
\begin{equation}
\sigma_{P_1} = \sqrt{2}\,\sigma\,,
\end{equation}
which is 0.0147 days (21.2 min).

This example used the maximum integrated flux value as the key feature to be identified in each cycle (period)
of the light curve.  This is, of course, an arbitrary choice.  One could instead choose any other unique feature
of the light curve, like
the minimum integrated flux value 
or the first (last) median flux crossing with a positive (negative) derivative.  The identification of these key features
in each cycle (period) of the light curve is easier with high signal-to-noise (S/N) observations.  Ambiguous key
feature identifications result in larger period errors than are found with high S/N observations.

\subsection{Two Cycles}

Let us now continue with another illustrative example.
Consider a 10-day time series of uninterrupted space-based 30-min observations
of a strictly periodic variable with a period $P\!=\!4.7$ days 
and a timing uncertainty of $\sigma\!=\!0.0104$ days.
Assume that the first peak of the light curve occurs during the first observation.
There will be three measurements with peak flux values at $t_1, t_2$ and $t_3$.

The estimated period from the single measurement with two periods is
\begin{equation}
P_2 = \frac{(t_3-t_1)}{2}\,,
\end{equation}
and the variance of $P_2$ is
\begin{equation}
\sigma^2_{P_2} 
= 
\left(\frac{1}{2}\right)^2
\left(
\sigma^2_{t_3} 
+ 
\sigma_{t_1}^2
\right)
= 
\left(\frac{1}{2}\right)^2
\left(2\sigma^2\right)
=
\frac{\sigma^2}{2}\,.
\end{equation}
The uncertainty of the period is 
\begin{equation}
\sigma_{P_2}= \frac{\sigma}{\sqrt{2}}\,,
\label{p2-2}
\end{equation}
which is 0.00735 days (10.3 min).

The estimated period from the two measurements with a single period is
\begin{equation}
2P_1 
= 
(t_3 - t_2)
+
(t_2-t_1) \,,
\end{equation}
and the variance of $P_1$ is
\begin{equation}
\sigma^2_{P_1} = 
\left(\frac{1}{2}\right)^2
\left(\sigma_{t_3}^2 +\sigma_{t_2}^2\right)
+
\left(\frac{1}{2}\right)^2
\left(\sigma_{t_2}^2 +\sigma_{t_1}^2\right)
=\sigma^2\,.
\end{equation}
The uncertainty of the period is thus
\begin{equation}
\sigma_{P_1} = \sigma\,,
\end{equation}
which is 15 min (0.0104 days).

If we combine the single double-period measurement with the two single-period measurement,
we find a period uncertainty of
\begin{equation}
\sigma_{P_{2+1}}
=
\frac{1}{\sqrt{2}}
\sqrt{
\sigma^2_{P_2}
+
\sigma^2_{P_1}\,,
}
\end{equation}
which
12.6 min (0.00900 days).
Note that this value is 2.3 min {\em{worse}} than the 10.3 min uncertainty
found using only the single measurement with two periods.
One must carefully use only measurements that improve knowledge:
adding lower-quality measurements can lead
to a worse result.

\subsection{Four Cycles}

Consider a 10-day time series of uninterrupted space-based 30-min observations
of a strictly periodic variable with a period $P\!=\!2.4$ days 
and a timing uncertainty of $\sigma\!=\!0.0104$ days.
Assume that the first peak of the light curve occurs during the first observation.
There will be five measurements with peak flux values at $t_1, t_2, t_3, t_4$, and $t_5$.

The estimated period from the single measurement with four periods is
\begin{equation}
P_4 = \frac{(t_5-t_1)}{4}\,,
\end{equation}
and the variance of $P_4$ is
\begin{equation}
\sigma^2_{P_4} 
= 
\left(\frac{1}{4}\right)^2
\left(\sigma_{t_5}^2 + \sigma_{t_1}^2\right)
= 
\left(\frac{1}{4}\right)^2
\left(2\sigma^2\right)
=
\frac{\sigma^2}{8}\,.
\end{equation}
The uncertainty of the period is 
\begin{equation}
\sigma_{P_4}= \frac{\sigma}{\sqrt{8}}
\label{p4-4}
\end{equation}
which is 5.29 min (0.00367 days).

The estimated period from the two measurements with three periods is
\begin{equation}
2P_3 = 
\frac{(t_5-t_2)}{3} 
+
\frac{(t_4-t_1)}{3} \,,
\end{equation}
and the variance of $P_3$ is
\begin{equation}
\sigma^2_{P_3} 
= 
\left(\frac{1}{6}\right)^2
\left(\sigma_{t_5}^2 + \sigma_{t_2}^2\right)
+
\left(\frac{1}{6}\right)^2
\left(\sigma_{t_4}^2 + \sigma_{t_1}^2\right)
=
\frac{\sigma^2}{9}\,.
\end{equation}
The uncertainty of the period is 
\begin{equation}
\sigma_{P_3}= \frac{\sigma}{3}
\label{p2}
\end{equation}
which is 5.00 min (0.00346 days). Since $\sigma_{P_3}$ is better than $\sigma_{P_4}$, we continue
this analysis.

If we combine the single quadruple-period measurement with the two triple-period measurements,
we find a period uncertainty of
\begin{equation}
\sigma_{P_{4+3}}
=
\frac{1}{\sqrt{2}}
\sqrt{
\sigma^2_{P_4}
+
\sigma^2_{P_3}
}
\label{p4p3}
\end{equation}
which
5.14 min (0.00357 days).
This value is 0.15 min better than the 5.29 min uncertainty
found using only the single measurement with four periods.  

The estimated period from the three measurements with two periods is
\begin{equation}
3P_2 = 
\frac{(t_5-t_3)}{2} 
+
\frac{(t_4-t_2)}{2} 
+
\frac{(t_3-t_1)}{2} \,,
\end{equation}
and the variance of $P_2$ is
\begin{eqnarray}
\sigma^2_{P_2} 
&=& 
\left(\frac{1}{6}\right)^2
\!\!\left(\sigma_{t_5}^2 + \sigma_{t_3}^2\right)
+
\left(\frac{1}{6}\right)^2
\!\!\left(\sigma_{t_4}^2 + \sigma_{t_2}^2\right)
+
\left(\frac{1}{6}\right)^2
\!\!\left(\sigma_{t_3}^2 + \sigma_{t_1}^2\right)
\nonumber
\\
&=&
\frac{\sigma^2}{6}\,.
\end{eqnarray}
The uncertainty of the period is 
\begin{equation}
\sigma_{P_2}= \frac{\sigma}{\sqrt{6}}
\label{p2}
\end{equation}
which is 6.11 min (0.00425 days).

If we combine the 
single quadruple-period measurement,
the two triple-period measurements,
and the three double-period measurements,
we find a period uncertainty of
\begin{equation}
\sigma_{P_{4+3+2}}
=
\frac{1}{\sqrt{3}}
\sqrt{
\sigma^2_{P_4}
+
\sigma^2_{P_3}
+
\sigma^2_{P_2}
}
\end{equation}
which
5.48 min (0.00381 days).
This value is 0.34 min worse than the 5.14 min uncertainty
found using the combination of the
single quadruple-period measurement 
and the two triple-period measurements.  

\subsection{One to Seven Cycles}

Using a new notation where $\sigma_n$ replaces $\sigma_{t_n}$, which was used
above to denote the uncertainty of the timing of the $n^{\rm{th}}$ observation,
we now show in 
Table\ \ref{tbl-one}
\ifdefined\wantmarginnotes
\marginpar{Tbl\ref{tbl-one}}
\fi
the expanded and reduced forms of the {\em{square}} of the period error 
estimates for 
a time series of uninterrupted space-based observations of a strictly periodic variable
with 1 to 7 complete cycles (periods).
The square of the period error estimates
was used in order to eliminate the need for square root symbols that would have otherwise
been required in many places within the table.
This table was typeset this way in order to help the reader see the underlying pattern
that is revealed when estimating period errors for observations with many complete
cycles.
The first column gives the value of $M$ which is defined as being
the maximum number of cycles that can occur in the time series.
The second and third columns, respectively, give the expanded and reduced forms
of the {\em{square}} of the period error estimate for a given value of $M$.
The reduction of the expanded forms assumes that
the timing uncertainty for all observations ($\sigma$) is the same for any given time series:
$
\sigma = \sigma_1 = \sigma_2 = \ldots = \sigma_{n-1} = \sigma_{n}\,.
$
We see from Table\ \ref{tbl-one} that 
the square of the period error estimate for an observation with 4 complete cycles is
\begin{equation}
\sigma^2_{M=4} 
= 
\frac{\sigma^2}{2}
\left[
\frac{1}{8}
+
\frac{1}{9}
\right]\,,
\end{equation}
which is the square of the value of Equation (17).

\subsection{The PEC Algorithm}

The pattern shown in Table\ \ref{tbl-one}
can be generalized to any number of complete cycles in a given time series
with the following simple algorithm:\\[0.5em]
Given three input parameters,\\
\hspace*{0.5\parindent}$L$ is the total length of the time series in days,\\
\hspace*{0.5\parindent}$\sigma$ 
is the timing uncertainty (one standard deviation) 
for a single flux value in days,\\
\hspace*{0.5\parindent}$P$ is the period of the variable in days,\\
and one derived parameter,\\
\hspace*{0.5\parindent}$M\equiv{\rm{int}}(L/P)$ 
which is the maximum number of periods that can occur in the time series, 
the square of the total measurement error for the period $P$ of a strictly periodic variable
can be estimated as
\begin{equation}
\sigma^2_{\rm{PEC}}
\equiv
\min(\,f(i;\sigma,M)\,){\rm{~for~}}i = 1, 2, 3,\,\dots, M,
\label{pec-1}
\end{equation}
where
\begin{equation}
f(i;\sigma,M) 
\equiv
\frac{1}{i}
\sum_{j=1}^i
\frac{2\sigma^2}{j\left(M-(j-1)\right)^2}
\,.
\label{pec-2}
\end{equation}\\[0.5em]
The name of the algorithm is PEC which stands for {\em{P\,}}eriod {\em{E\,}}rror {\em{C\,}}alculator.
With time series that span many cycles (periods), computing the PEC algorithm by hand quickly becomes tedious
as the number of terms in the optimal solution
can easily have hundreds to thousands of terms ---
which is typically the case if one is analyzing 90-day long cadence
light curves from the Kepler Mission. 
Fortunately, the  PEC 
algorithm can be efficiently programmed using just a few lines of C computer language code\footnote{
A C implementation of the PEC algorithm ({\tt\small{pec.c}}) is available at the following website:\newline
{\tt\footnotesize{http://www.noao.edu/staff/mighell/PEC}}~.
}
\citep{PECascl}.

\begin{figure}
\center{
\includegraphics[width=0.90\textwidth]{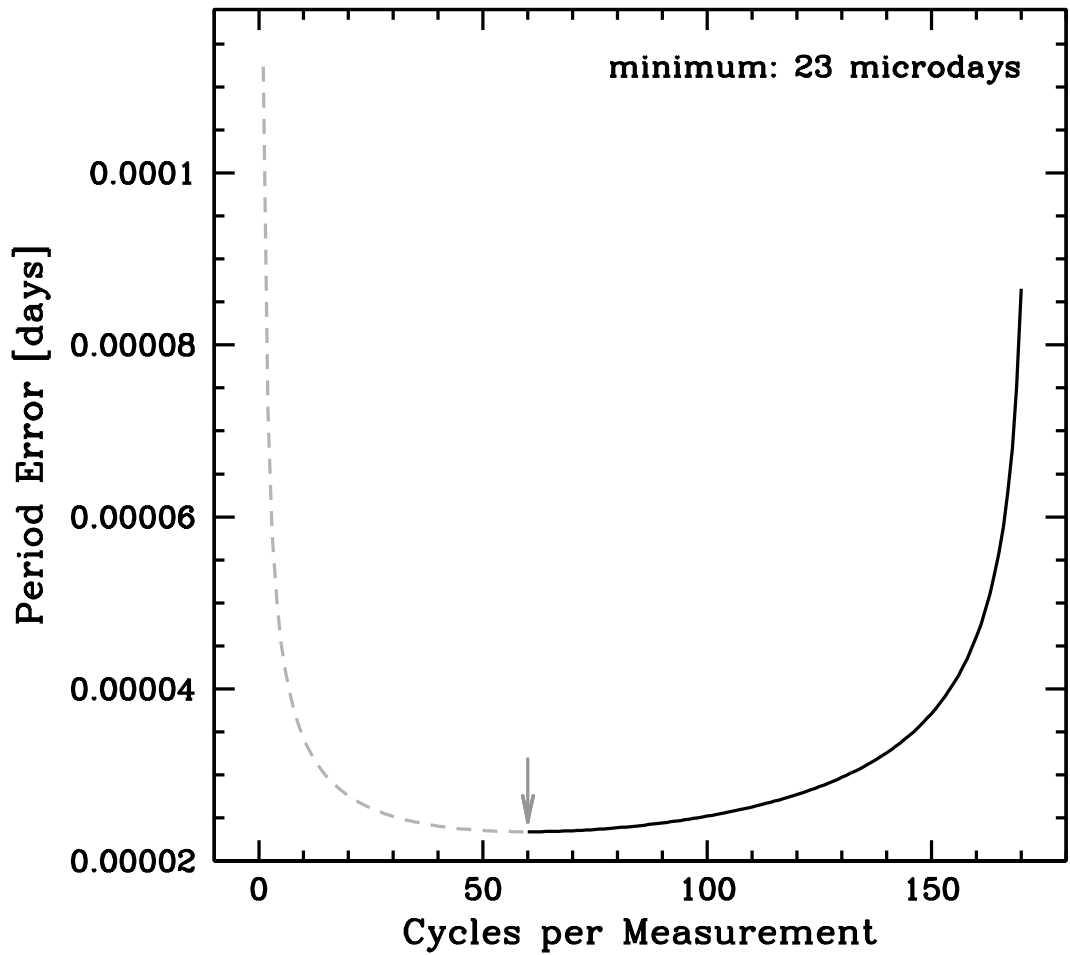}
\caption{
PEC algorithm analysis of a 90-day time series of a strictly periodic variable with a period $P$$=$$0.5274$ days
and a timing uncertainty (half-bin width) of $\sigma$$=$$0.0104$ days (15 min).
The arrow points to the minimum of the curve which is 23 microdays at 60 cycles per measurement.
\newline
\label{PEC}
}
}
\end{figure}

Figure \ref{PEC} 
\ifdefined\wantmarginnotes
\marginpar{Fig\ref{PEC}}
\fi
shows the PEC algorithm 
analysis of a 90-day time series observation of a strictly periodic variable with a period $P$$=$$0.5274$ days
and a timing uncertainty (half-bin width) of $\sigma\!=\!0.0104$ days (15 min).
The solid curve shows the progress of the PEC algorithm as it starts with one 170-cycle measurement and progresses
to the left
until it includes 111 60-cycle measurements 
where upon it finds the optimal (minimum) period error estimate of 23 microdays which is based on 6216
individual measurements each of which has a minimum of 60-cycles per measurement.
The dashed curve shows how the solution becomes progressively {\em{worse}} 
if one chooses to ignore the stopping condition of the minimum function in Equation (\ref{pec-1}) by including more 
and more lower-quality measurements with fewer than 
60 cycles per measurement.
In this example, 
the stopping point of 60-cycles measurements is independent of the value of the timing uncertainty, $\sigma$,
because $\sigma_{\rm{PEC}}$ is linearly proportional to the value of the timing uncertainty 
(see Equations (\ref{pec-1}) and (\ref{pec-2}));
reducing $\sigma$ by a factor of 100 from 15 min to 9 s (0.000104 days) 
gives a period error estimate that is 100 times smaller than before: 0.23 microdays.

The use of the minimum function in Equation (\ref{pec-1}) as a stopping condition 
is required in order to obtain the optimal (minimum) value for the period error estimate.
Note that the value of the summation elements
within the square brackets shown in the
third column of Table \ref{tbl-one}
successively decrease.
Once the value of the summation elements start increasing, the value of the summation will be
greater than the minimum value.
Consider 
the case where we only include one single $M$-cycle measurement of an observation that is $M$ cycles long
($i=1$):
\begin{equation}
f(1;\sigma,M) 
=
\frac{1}{1}
\sum_{j=1}^1
\frac{2\sigma^2}{j\left(M-(j-1)\right)^2}
=
\frac{2\sigma^2}{M^2}\,.
\end{equation}
This gives 
period error estimates of 
$\sigma / \sqrt{2}$ 
and
$\sigma / \sqrt{8}$,
respectively,
for a 2-cycle and a 4-cycle observation,
as previously shown in 
Equations 
(\ref{p2-2})
and
(\ref{p4-4})\,.
Now consider the case where {\em{all}} possible measurements are made ($i=M$, where $M>1$):
\begin{equation}
f(M;\sigma,M) 
=
\frac{1}{M}
\sum_{j=1}^M
\frac{2\sigma^2}{j\left(M-(j-1)\right)^2}
=
\frac{1}{M}
\left[
\frac{2\sigma^2}{M^2}
+
\ldots
+
\frac{2\sigma^2}{M}
\right]\,.
\label{fmm}
\end{equation}
This gives a period error estimate of 
$\sim\!0.8660\,\sigma$ 
for a 2-cycle observation,
which is larger than the optimal period estimate of
$\sigma_{\rm{PEC}} \approx 0.7071\,\sigma$ (see Table \ref{tbl-one} and Equation (\ref{p2-2})).
This gives a period error estimate of 
$\sim\!0.4751\,\sigma$ 
for a 4-cycle observation,
which is larger than the optimal period estimate of
$\sigma_{\rm{PEC}} \approx 0.3436\,\sigma$ (see Table \ref{tbl-one} and Equation (\ref{p4p3})).
The fact that the last summation element ($2\sigma^2\!/M$) in Equation (\ref{fmm})
is larger than the first summation element ($2\sigma^2/M^2$) indicates that
one should never include (except in the simplest case of $M$=1)
measurements of 1-cycle observations when
determining period error estimates
because doing so will produce a worse period error estimate
than if one just measured a single $M$-cycle observation.
That is why no 1-cycle observations 
are included in Table \ref{tbl-one}
except when $M$=1\,. 
The stopping condition of the minimum function in Equation (\ref{pec-1}) ensures
that only measurements which improve the period error estimate are used
in the computation of $\sigma_{\rm{PEC}}$.

\section{PERIOD ERROR ESTIMATES FOR THE KEBC}

We have shown above how period error estimates of eclipsing binaries
can be determined with many days of idealized\footnote{The issue of using time series observations 
that are not ideal will be discussed below.} uninterrupted (space based) observations. 
Estimating the period errors of eclipsing binary systems based
on many days of real Kepler Mission long-cadence observations
is slightly more complicated.

The above examples assume high signal-to-noise observations in that the expectation is that the 
integrated peak flux is observed in the cadence when the peak flux actually occurred.  
This assumption should be appropriate for most of the Kepler light curves studied in this article.
However, with low S/N observations
photon noise can cause ``bin-hopping''
where the integrated peak flux is found in {\em{neighboring}} cadences (measurements).
This phenomenon with low S/N observations can conservatively be accounted for 
by tripling the timing uncertainty of high S/N observations (0.5-bin width to 1.5-bin width)
which implies that the true peak may actually occur one cadence {\em{before}} or {\em{after}} the cadence 
with the maximum (peak) flux value.
Period errors scale with the timing uncertainty and so 
the downside of this ``fix'' is that the resultant period error will be three times worse
than in the high S/N case. 

The total integration time for a normal Kepler long cadence observation is about 1765.5~s ($\,=\!29.4$ min)
\citep[see Figure 20 of][]{KeplerDataCharacteristicsHandbookMarch2012}
which rounds up to the commonly used approximation of 30 min.  An integration time of 30 min is 
an excellent approximation of the 29.4 min (a 2\% error) integration time of real Kepler long cadence observations.

The quoted times\footnote{ 
The barycentric times currently reported in the {\tt{TIME}} columns and the headers of all Kepler data products
are currently wrong by more than a minute. The reported times
can be corrected to the TDB (Barycentric Dynamical Time) system by adding 66.184 s to the reported 
barycentric times for all cadence numbers less than or equal to 57139 in LC (long cadence).
For times after this cadence, one needs to add 67.184 to account for the recent leap second at
UTC 2012-06-30 23:59:60.
Except for the addition of one leap second in Q14, the reported times are internally consistent and this
error is only apparent when comparing Kepler times to other observations with timing accuracies better than a 
couple of minutes.
The relativistic correction between the UTC and the TDB systems,
which is of order 1.6 ms and is significantly less than the 50 ms precision of the Kepler clock,
is {\em{not}} accounted for with this simple additive correction
\citep{KeplerDataReleaseNotesQ14}.
}
for any Kepler cadence observation are believed to be accurate to within $\pm$50 ms
\citep{Koch+2010,KeplerDataCharacteristicsHandbookMarch2012,KeplerArchiveManualV4},
but this has not yet been tested with astrophysical data
\citep{KeplerDataCharacteristicsHandbookMarch2012}.
The Kepler Data Characteristics Handbook
\citep{KeplerDataCharacteristicsHandbookMarch2012} warns users who require
temporal accuracy better than one minute that they should read Section 6 (Time and Time Stamps) carefully.
The 50 ms timing accuracy for Kepler was a requirement
that was levied on Ball Aerospace (the Kepler Mission prime contractor)
in the top level Kepler Science Requirement Document and the Mission Requirement Document, both written
in 2002, but are not generally available (Douglas Caldwell, private communication 2013).
The 50 ms timing uncertainty was reported 
in Table 1 of Koch's Kepler Mission overview ApJ Letter \citep{Koch+2010}.
Since the actual timing accuracy of the Kepler spacecraft has yet to be tested with astrophysical data,
researchers should use the $\pm$50 ms timing
accuracy requirement as a {{systematic}} error rather than a random error which can be beaten down
with many measurements.
Until the timing accuracy of the Kepler spacecraft is proven to be better than 50 ms,
no more than six decimal places should be reported for any period measured in days
from Kepler observations
of periodic variables in the Kepler Field.

The timing uncertainty of a light curve peak found in a Kepler long cadence observation of a periodic variable
is {\em{not}} the $\pm$50 ms timing uncertainty of the Kepler clock.  
Since Kepler long cadence observations are
long integrations --- all that one really knows for sure is the peak flux occurred {\em{sometime}}
during the $\sim$1765.5 s of exposure time.
Using a one-sixth bin width as a timing uncertainty ($1\,\sigma$) will ensure that the peak actually did occur within the
long cadence observation with a $\sim\,$99.7\% ($\pm$3$\sigma$) 
probability.  Unfortunately, such a timing uncertainty gives too high
a probability of the peak occurred near the middle of the observation rather than near the 
beginning or end of
the observation.
Using a half-bin width as a timing uncertainty ($1\,\sigma$) is a more realistic approximation of the
timing uncertainty --- even though the use of that value implies that the peak occurs within the observation 
only $\sim\,$68.3\% ($\pm$1$\sigma$) of the time rather than the expected 100\% probability 
(assuming high signal-to-noise observations).
It is prudent to be conservative rather than to add a hidden bias to a statistical analysis.

\begin{figure}
\center{
\includegraphics[width=0.90\textwidth]{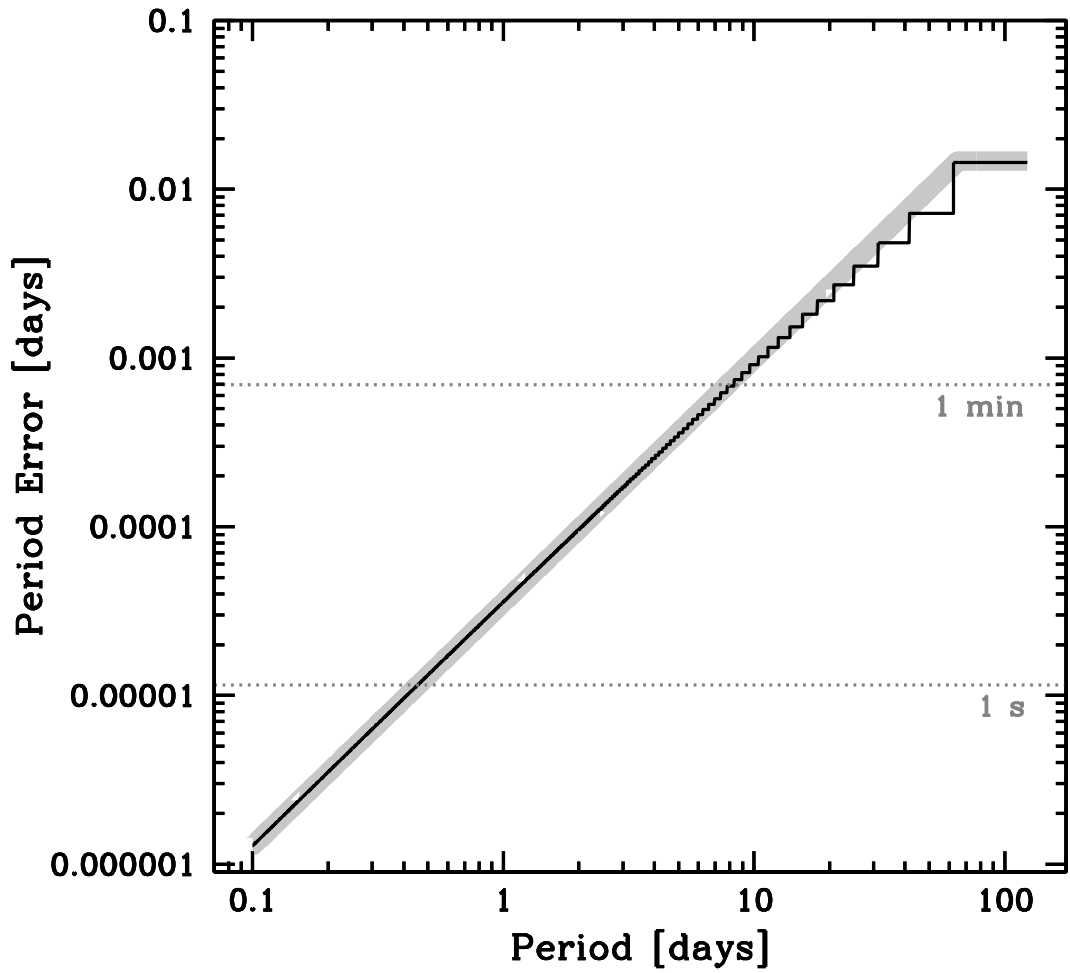}
\caption{
The jagged curve shows the results of the 
PEC algorithm analysis 
for the period errors of the Kepler Eclipsing Binary Catalog.
The gray curve shows the analytical model of the PEC results.
\newline
\label{graph}
}
}
\end{figure}

The predicted period errors of the Kepler Eclipsing Binary Catalog
are shown as the jagged curve of Figure \ref{graph}.
\ifdefined\wantmarginnotes
\marginpar{Fig\ref{graph}}
\fi
The curve is the
PEC algorithm analysis of an uninterrupted 125-day time series observations of
strictly periodic variables with periods ranging from 0.1 days to 125 days
and a timing uncertainty (half-bin width) of $\sigma\!=\!0.0102$ days (14.7 min).
The curve is jagged for longer periods because there are only so many integer number
of long cycles that can be fit into a 125-day time series. The extreme example of this
is that it is only possible to get one single period measurement 
of a periodic variable with a period between 62.5 and 125 days
during a single 125-day time series and the
predicted period errors are all the same:
$0.0144$ 
$(=\sqrt{2}\,\sigma)$ 
days (20.8 min). 

The period errors estimated by the PEC algorithm are {\em{optimistic}} compared to real Kepler observations.
While the vast majority of Kepler observations are excellent, some fraction of the observations are problematic.
The Kepler pipeline has matured sufficiently since the start of the Kepler Mission
such that users of the Kepler Data Archive can now confidently reject
most ``bad'' data by simply not using observations with {\tt{SAP\_QUALITY}} values greater than zero.

The PEC algorithm assumes continuous observations --- which may or may not describe any given Kepler observation.
Long observations of a target will have bad observations ({\tt{SAP\_QUALITY}} $> 0$) or may contain
a gap as large as $\sim\,$24 hours when no observations were obtained because
the Kepler spacecraft moved off the Kepler Field in order to position itself to download
all the observations obtained during the previous month.

Missing data is missing information. If many entire cycles are missing from the time series
(e.g., gaps due to data download sessions),
then period errors estimated from such patchy data may well be
greater than they would have been with 100\% coverage. 
However, if the data loss occurs cleanly between
the eclipse and transit of a long-period detached eclipsing binary, then the period
error estimate may be as good as what could have been determined without any data loss.

A simple model that
approximates the {{upper envelope}} of the jagged curve of the PEC algorithm predicted period errors
can be easily developed.
The PEC algorithm predicts period errors 
of 1.2861 and 35.622 microdays
for two variables with periods of $0.1$ and $1$ days, respectively.
After converting these predicted errors into common (base 10) logarithm values and doing some simple arithmetic,
the following simple
model estimates the period errors for the Kepler Eclipsing Binary Catalog:
\begin{equation}
\sigma_{\rm{KEBC}}
\approx
\left\{
\begin{array}{l}     
    10^{-5.8908 + 1.4425\,\left( \log(P_{\rm{KEPC}}) + 1\right)}{\rm{~days}}\,,\\ 
   {\rm{\hspace*{\parindent}if~}} P_{\rm{KEBC}} < 62.5 {\rm{~days}}\,;\\
   0.0144 {\rm{~days\,,~}}\\
   {\rm{\hspace*{\parindent}if~}} P_{\rm{KEBC}} \geq 62.5 {\rm{~days;}}
\end{array}\right.
\label{eq:kebc}
\end{equation}
where $P_{\rm{KEBC}}$ is the period of the variable in the KEPC in units of days.
This model is shown as the thick gray curve in Figure \ref{graph}.
The PEC algorithm predicts a period error of 
180 microdays for a strictly periodic variable with a period $P\!=\!3.17$ days;
Equation (\ref{eq:kebc}) predicts a period error of 190 microdays (an error of 5.6\% greater than
the PEC algorithm estimate).
The PEC algorithm predicts a period error of 
910 microdays for a period $P\!=\!10$ days;
the model predicts a period error of 990 microdays (an error of 8.8\%).
The PEC algorithm predicts a period error of 
7.2 millidays for a period $P\!=\!41.69$ days;
the model predicts a period error of 7.7 millidays (an error of 6.9\%).

So far, we have only considered theoretical predictions for period errors.
How well do these predicted period error estimates 
compare to real period error measurements based on Kepler long cadence observations
of two semidetached and five detached eclipsing binaries and an RR Lyrae variable star?

The NASA Exoplanet Archive Periodogram Service\footnote{The 
website of the simple upload version of the NASA Exoplanet Archive Periodogram Service is\newline
{\tt{http://exoplanetarchive.ipac.caltech.edu/cgi-bin/Periodogram/nph-simpleupload}}~.}
is a U.S.\ Virtual Astronomical Observatory (VAO) web service \citep{VAO2011annualreport,VAO2012annualreport}
that searches for periodic signals in exoplanet observations from the NASA Kepler
and the ESA CoRoT astrophysical missions or in user-provided observations of time and flux of variable astronomical objects.
The Periodogram Service returns periodograms of time series data.
Users can download the derived periodograms
as well as the phased light curves for the five most significant periods determined from the data.
Unfortunately, no period error estimates are provided for any significant periods.

At the beginning of every Kepler quarter since Q2, the Kepler spacecraft rotates 90$\arcdeg$
in order to optimize the orientation of the solar panels on spacecraft
with respect to the Sun.  As a consequence of this roll,
when the spacecraft resumes its normal operations, light from a given target now falls on a 
different CCD than was used in the previous quarter.  Individual Kepler observations are not flux
calibrated in an absolute sense and so there can be considerable jumps (up and down) in the
measured flux of a given target from quarter to quarter
--- even if there is no evidence for variability in the target
{{and}} the same aperture is used.
If one wishes to analyze multiple quarters of Kepler observations simultaneously, it is necessary
to ``normalize'' the data from different quarters to be approximately on the same flux scale.
For example, for detached eclipsing binaries with long periods, the measured flux between eclipses
should be the same --- from quarter to quarter.  
If one plots the light curve of a given target 
(e.g., {\tt{PDCSAP\_FLUX}} vs.\ {\tt{TIME}}) 
for multiple Kepler quarters, one sees that the maximum and minimum flux values can
vary a lot between quarters. The normalization process across multi-quarter observations
can be quite complicated
(e.g., see \S 4.2 (Data Detrending) of \citet{Slawson+2011})
and is beyond the scope of this article.
The Periodogram Service has recently implemented an option to normalize Kepler
light curves for all public data for a given Kepler target. 
The Periodogram Service generally does
a good job of normalizing flux values across many Kepler quarters.  
While the current normalization
algorithm is not perfect, it is a good step in the right direction. 

The Periodogram Service gives the user a choice of three algorithms for computing periodograms from light curve data:
(1) Lomb-Scargle \citep{lomb-scargle},
(2) ``BLS'' (Box-fitting Least Squares) \citep{bls},
and 
(3) Plavchan \citep{plavchan}.
The Lomb-Scargle algorithm is the default method used by the Periodogram Service.
A description of the implementation details of these three algorithms 
may be found on the Algorithm Documentation webpage
of the Periodogram Service\footnote{Periodogram Service algorithm documentation webpage:\newline
http://exoplanetarchive.ipac.caltech.edu/applications/Periodogram/docs/Algorithms.html\ .
}.

While the Periodogram Service currently does not directly provide period error estimates for the
periods it finds as peaks in the power of computed periodograms, a user of the Periodogram
Service can measure the period error of a given period by using the method described below.
A major drawback of this method is 
that it is both
labor intensive for the user 
and 
computationally intensive for the 128-core computer cluster used by the Periodogram Service
(seconds to minutes per eclipsing binary system, depending on the current user load on the cluster).

Over a narrow range of periods near the best period estimate, periodograms of eclipsing
binaries are typically bell-shaped (non-Gaussian) curves.
A good estimate of the uncertainty of a period measurement is 
the Half-Width-at-Half-Maximum (HWHM) of 
the periodogram peak. 

\begin{figure}
\center{
\includegraphics[width=0.90\textwidth]{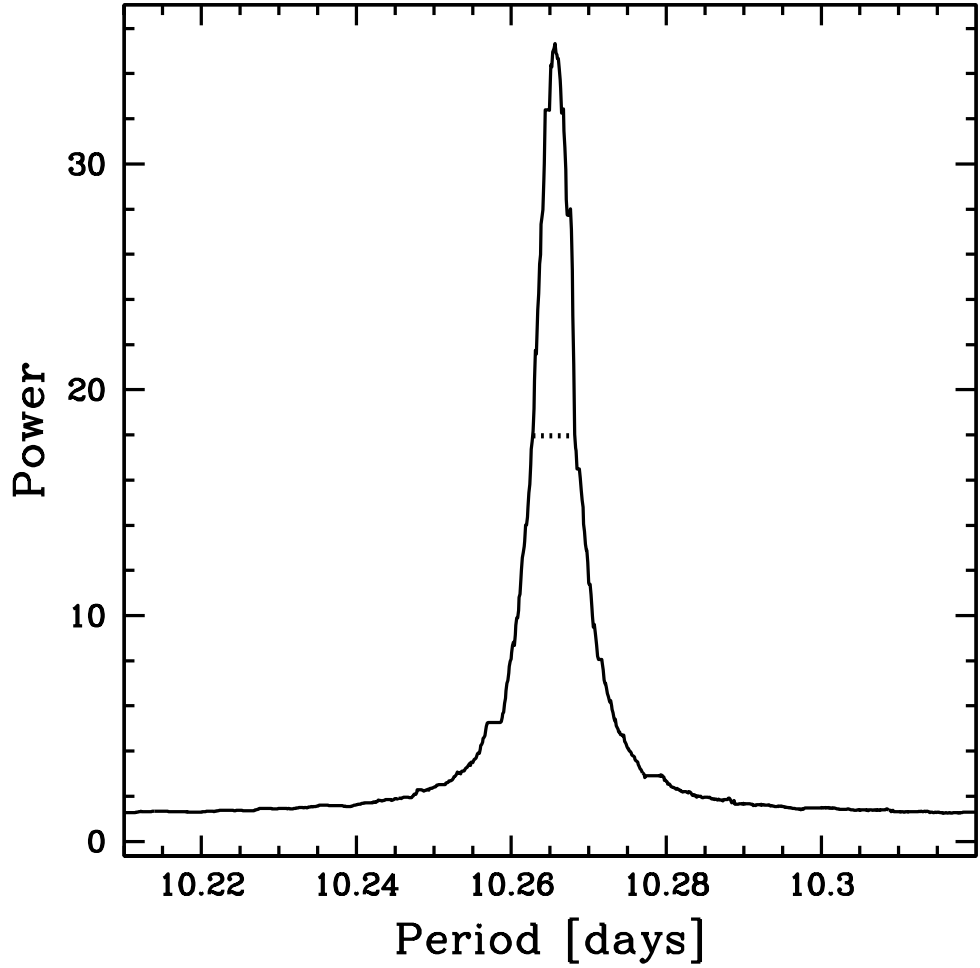}
\caption{
Periodogram of the Plavchan algorithm analysis of normalized Kepler Q0, Q1, and Q2 observations of the 
detached eclipsing binary KID 3120320.
The dotted line is the Full-Width-at-Half-Maximum (FWHM) which is 5.6 millidays wide.
The measured period for KID 3120320 is 10.2656(28) days.
\newline
\label{KID3120320}
}
}
\end{figure}

The peak of the periodogram (maximum power), produced by the Plavchan algorithm
analyzing normalized Kepler observations from quarters Q0 through Q2,
of the detached eclipsing binary KID 3120320 occurs at 10.2656 days
and the Full-Width-at-Half-Maximum (FWHM) of the periodogram peak is 5.6 millidays wide
(see Figure \ref{KID3120320}).
\ifdefined\wantmarginnotes
\marginpar{Fig\ref{KID3120320}}
\fi
This periodogram was computed 
over a much narrower period range (9.75--10.75 days) and
at a much higher resolution (430$\times$) in frequency-space 
and with a much higher resolution phase-smoothing box size (30$\times$) than is
normally provided by default by
the Periodogram Service using the Plavchan method: 
{\tt{fixed df step size}} of 0.000001 and a {\tt{phase-smoothing box size}}
of 0.002\,.  All of these modifications to the default parameters must be done by
hand by the user because, as of now, such changes can not be done remotely
using a computer script.

The uncertainty of the period estimate is approximately the HWHM: 2.8 millidays.
The period estimate of KID 3120320 is thus
$P\!=\!10.2656(28)$ days,
which agrees exactly with the Kepler Eclipsing Binary Catalog estimate of 10.265600 days --- 
but the last 2 digits of the KEBC period are probably not significant.  

The Plavchan algorithm was found to produce more precise periods
and period error measurements of Kepler observations of eclipsing binaries than
the Lomb-Scargle and BLS algorithms.  With the same data for KID 3120320,
the
Lomb-Scargle algorithm gave a period of 10.20(31) days
while the BLS algorithm gave a period of 10.24(11) days.
In this case, the Plavchan algorithm produced two more significant digits of precision for the period measurement
as compared with the Lomb-Scargle and the BLS algorithms.

The top two sections of
Table \ref{measurements} 
\ifdefined\wantmarginnotes
\marginpar{Tbl\ref{measurements}}
\fi
gives the period and period error measurements of 7 eclipsing binaries
in the Kepler Eclipsing Binary Catalog with periods ranging from $\sim\,$0.5 to $\sim\,$50 days.
The first column gives the Kepler quarters that were used in the analysis; Kepler quarters Q0 through 
Q2
were normally used (as was done with the KEBC) but in the case of KID 9540450 there were no Q0 observations.
The second column gives the unique Kepler Identification number (KID) for each target.
The third column gives the type of eclipsing binary system of the target; 
the two shortest period systems are semi-detached eclipsing binaries and the rest are detached eclipsing binaries.
The fourth column gives the brightness the target in Kepler magnitudes (KEPMAG) as given
in the Kepler Input Catalog (KIC) \citep{Brown+2011}.
The fifth column gives the period of the target as given in the KEBC.
The sixth column gives the measured period and period error as determined using the Periodogram Service.
The last column gives the period error estimate of the PEC algorithm (Equation (\ref{pec-1}))
for a uninterrupted 125-day time series
and a timing uncertainty (half-bin width) of 0.0102 days 
$\left[=\!\left(29.4{\rm{~min}}\right)/2\right]$.

The measured period errors for the top section of Table \ref{measurements}
for the eclipsing binary systems KID 11560447, 10858720, 9873869, 3120320, and 9172506
are, respectively, 
39, 63, 570 microdays, and 2.8, 19 millidays (shown as diamonds in Figure \ref{graph2}).
\ifdefined\wantmarginnotes
\marginpar{Fig\ref{graph2}}
\fi
These measured period 
errors are larger than the PEC algorithm estimate by factors of 
2.8, 
1.9, 
1.7, 
3.1, 
and 2.6, respectively.

The middle section of Table \ref{measurements} investigates how the measured period error changes
as a function of brightness of two eclipsing binaries with nearly identical periods ($P\!\approx\!2.16$ days).
The detached eclipsing binary systems KID 9641031 and 9540450 have Kepler magnitudes of 9.177 and
14.146 mag with measured period errors of 
0.13 and 0.26 millidays
respectively.
Bright eclipsing binary systems have smaller period errors than faint systems with the same period.

The gray region in Figure \ref{graph2} shows where accurate and precise period error measurements of
systems in the Kepler Eclipsing Binary Catalog are expected to be found.
The bottom limit of the gray region is the jagged curve shown in Figure \ref{graph}; high S/N observations
are expected to be found near the bottom limit.
The top limit of the gray region is 3 times the jagged curve shown in Figure \ref{graph}; low S/N observations
are expected to be found near the top limit.
Note that the measured period errors of the seven eclipsing binaries in Table \ref{measurements}
lie within or very near the gray region in Figure \ref{graph2}. 
As expected, the bright eclipsing binary KID 9641031 ($P$=2.17816(13) days, 9.177 mag) lies near the
bottom limit of the gray region while the faint binary KID 9540450 ($P$=2.15472(26) days, 14.146 mag)
lies just slightly above the top limit of the gray region 
($3.1\sigma_{\rm{PEC}}$ instead of $3\sigma_{\rm{PEC}}$).

While very bright eclipsing binary systems in the Kepler Eclipsing Binary Catalog can nearly achieve the
period errors predicted by the PEC algorithm (see Figure \ref{graph2}),
for most eclipsing binary systems in the KEBC,
the period errors predicted by the PEC algorithm are likely to be the {\em{lower limit}} of the true period errors.
In other words,
most periods in the KEBC are likely to have uncertainties that are larger than the period error estimated 
using the PEC algorithm.
Nevertheless, the PEC algorithm (Equation (\ref{pec-1})) or its approximation for the 
KEBC (Equation (\ref{eq:kebc})) can
provide useful limits to the true precision of periods in the KEBC.

\section{DISCUSSION}

The information content of the phased light curve of long-period detached eclipsing binaries is low ---
only observations obtained during a transit (primary eclipse) or secondary eclipse offer useful information about the period.

The PEC algorithm predicts period errors based only on a single feature (the peak/minimum value) of each cycle
of a strictly periodic variable.
If the phased light curve of a strictly periodic variable is morphologically complicated, then the period error
estimated by the PEC algorithm can be reduced by a factor of $\sqrt{n}$ 
if $n$ unique
features can be robustly identified in each cycle of observation.
For example, if it is possibly to unambiguously identify both the maximum and minimum of a
light curve during each cycle of observation, then the PEC algorithm period estimate can be reduced
by a factor of $\sqrt{2}$.  

Theoretically, the more unique features one can measure,  
the better the improvement will be --- but one quickly encounters a situation of diminishing returns.
Suppose we analyze Kepler long cadence observations of a RR Lryae variable with a period of 0.5 days.
Phased light curves of RR Lyrae are much richer morphologically than phased light curves of eclipsing
binaries. There will be 24 Kepler long cadence observations per cycle (period).  
An improvement of $\sqrt{24}$ over the PEC
algorithm period error estimate would be the best possible case. But that would 
require a unique ``feature'' to be present in {{every single Kepler long cadence
observation}}.

The last section of Table \ref{measurements} investigates how the measured period error 
of FN Lyrae (KID 6936115), a non-Blashko ab-type RR Lyrae \citep{Nemec+2011}, 
compares to the PEC algorithm period error estimates.
Using the Periodogram Service to analyze Kepler quarter Q2 observations, we determined that 
the period of FN Lyr 
is 0.5273989 days with an uncertainty of 4.4 microdays.
The PEC algorithm gives a period error estimate of 23 microdays for an uninterrupted 88.9-day 
\citep{KeplerDataReleaseNotesQ0toQ4} time series
and a timing uncertainty (half-bin width) of 0.0102 days.
The measured period error is a factor of 5.2 times better than the PEC algorithm estimate
and is slightly better than the estimated ``best possible case'': 4.6 microdays
($\approx$23 microdays$/\sqrt{24}$) assuming that all cadences have a unique feature.
Using the Periodogram Service to analyze seven normalized Kepler quarters of observations (Q2 through Q8), 
the period of FN Lyr was determined to be 0.5273989 days with an uncertainty of 1.4 microdays.
Although the period estimate is the same as before, the uncertainty of the measurement
has been reduced by a factor of 3.1.
The PEC algorithm gives a period error estimate of 1.4 microdays for an uninterrupted 608.9-day 
\citep{KeplerDataReleaseNotesQ0toQ4,KeplerDataReleaseNotesQ5toQ8}
time series
and a timing uncertainty (half-bin width) of 0.0102 days.
Note that the measured period error using 7 Kepler quarters (Q2--Q8) agrees {{exactly}} with the PEC algorithm estimate.
Figure \ref{graph3} 
\ifdefined\wantmarginnotes
\marginpar{Fig\ref{graph3}}
\fi
may prove useful for planning future Kepler Mission observations of eclipsing binaries
and other periodic variables
in the Kepler Field.

\begin{figure}
\center{
\includegraphics[width=0.90\textwidth]{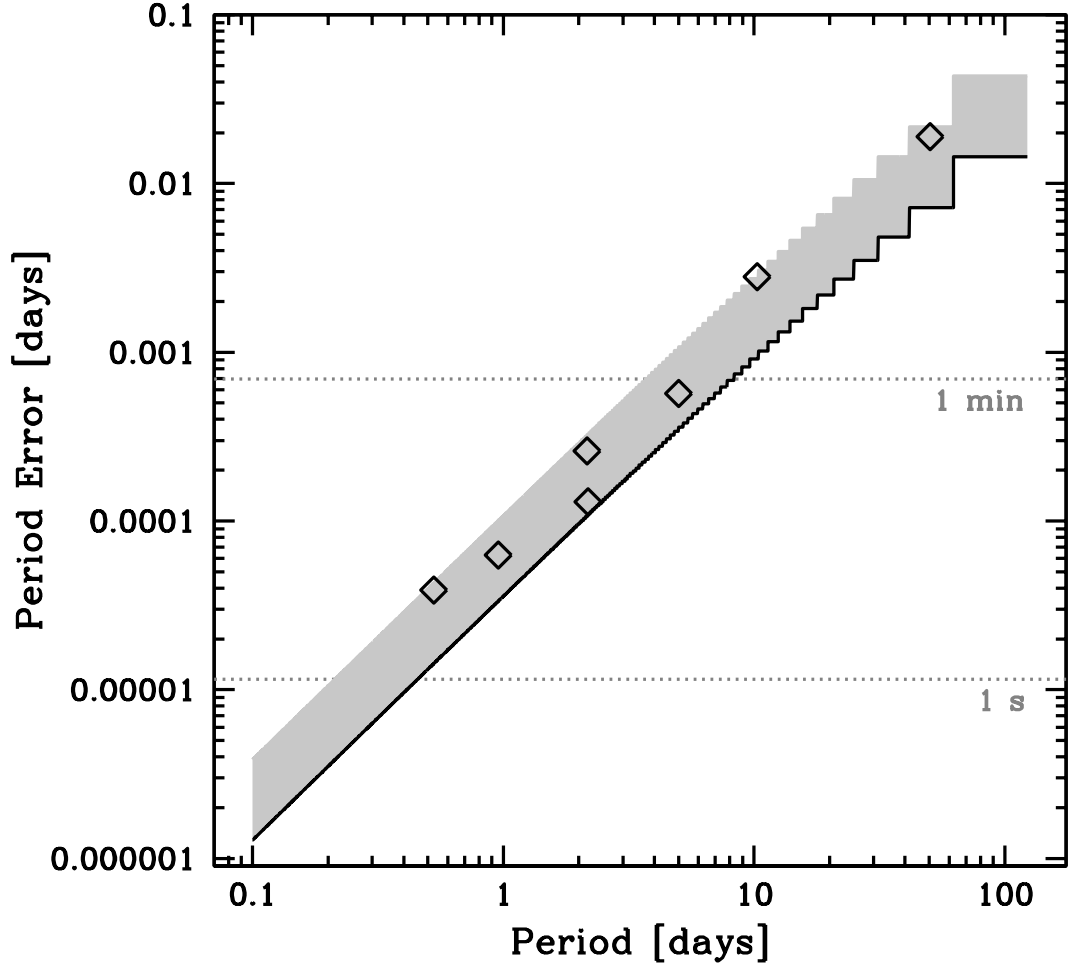}
\caption{
The jagged curve shows the results of the 
PEC algorithm analysis 
for the period errors of the Kepler Eclipsing Binary Catalog
(same as Figure\ \ref{graph}).
The diamonds show the measured period errors of the seven eclipsing binaries given in Table \ref{measurements}.
The gray region shows where accurate and precise period error measurements are expected to be found.
\newline
\label{graph2}
}
}
\end{figure}

\begin{figure}
\center{
\includegraphics[width=0.90\textwidth]{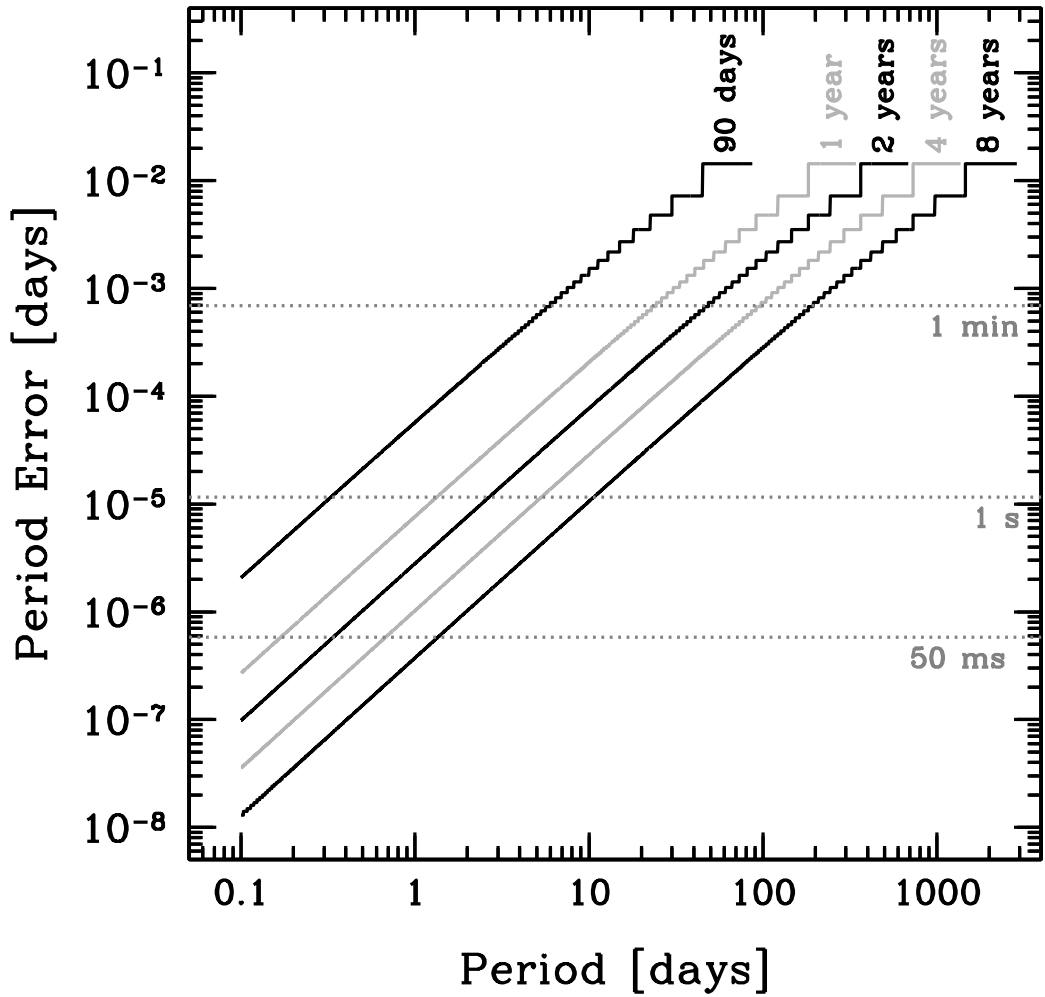}
\caption{
Period errors determined with the
PEC algorithm for Kepler long cadence observations with
durations of
90 days, 1 year, 2 years, 4 years, and 8 years.
\newline
\label{graph3}
}
}
\end{figure}

The PEC algorithm was devised to provide a reasonable estimate of the periods published in the 
Kepler Eclipsing Binary Catalog.  Figure \ref{graph2} shows that this goal has been accomplished.
The PEC algorithm was used to estimate the period errors of seven eclipsing binaries from the KEBC 
(see Table \ref{measurements}).  All seven error estimates were found in or near the gray region of
Figure \ref{graph2} which has lower and upper limits set by the PEC algorithm assuming, respectively,
high and low S/N measurements of the Kepler light curves.
While not perfect, the PEC algorithm is certainly a lot better than what we have now with regards
to KEBC period error estimates --- nothing at all.  

The big advantage of the PEC algorithm is that
it is a quick-look method for the determination of period error estimates for long time series
observations of eclipsing binaries.
It is fast and does not require any detailed analysis of actual light curve data; all that is required are 3 numbers
in units of days:
(1) the estimated period of the strictly periodic variable, 
(2) the timing uncertainty for a single measurement (0.0102 days for a Kepler long cadence observation), 
and 
(3) the total length of the continuous observation.
The 
curators of the NASA Exoplanet Archive Periodogram Service are encouraged to enhance the
Periodogram Service
by adding PEC algorithm period error
estimates when it reports likely periods based on periodogram analysis of Kepler light curves
of periodic variables.

In our description of the PEC algorithm we have referred to strictly periodic eclipsing binary systems.
Some eclipsing binaries, however, undergo period changes.  Sometimes those changes are due to 
the gravitational effects of a third body \citep[e.g.,][]{Hoffman+2006}.
Sometimes those changes can be due to mass transfers
in compact eclipsing binary systems \citep[e.g.,][]{Zhu+2012}.
Starspots as well as photometric noise can also cause subtle timing changes
of light curve minima \citep{Kalimeris+2002}.
Period changes are usually detected using $O-C$ diagrams where
the difference between observed time, $O$, of a light curve extremum (transit or eclipse) and the 
computed (predicted) time, $C$, is plotted using an ephemeris model.  Eclipsing binary systems with
significant $O-C$ differences are, by definition, not strictly periodic and thus the true period error
estimate for these systems will be larger than that predicted by the PEC algorithm.
Finding significant $O-C$ differences generally requires a detailed inspection of the actual light curve.
Now if a non-strictly periodic eclipsing binary system were to be continuously observed by Kepler
over the course of many years, it might be possible to {\em{detect}} such systems by comparing the
differences of orbital periods derived, from say yearly baselines, to the PEC period error estimate
for one year of Kepler observations.
If the $O-C$ differences only become significant over a period of many years or decades,
then the detection of such non-strictly periodic systems by Kepler 
using the PEC period error estimate
might not be possible due
to its currently envisioned maximum lifetime of 7.5 years for the Kepler Mission.

The neural networks currently
being used by the EBAI method could be improved to include the computation of confidence
intervals for the eclipsing binary parameters determined by its neural networks.
That might require a complete rewrite of the current neural network software framework used by the EBAI method,
which would probably be a time-consuming expensive exercise.  Although the end result may well be worth
the effort, especially in the era of a fully operational LSST (Large Synoptic Survey Telescope) project,
acquiring the necessary resources to make such significant changes might be challenging.
Alternatively, such an effort could possibly be cast as an astrophysics Ph.D. dissertation project 
with a major focus on applied
information systems research.

\cite{Nemec+2011} report that the period of FN Lyr is 0.527398471(4) days based on Kepler Q0 through Q5
observations.  That period error estimate is just 4 nanodays --- which is a factor of 145 smaller than the 
50 ms (579 nanodays) timing accuracy of the Kepler spacecraft.
``Typically, the periods derived using the Q0-Q5 LC data 
alone are accurate to 1--$2 \times 10^{-7}$ days" \cite{Nemec+2011}\,.
Assuming a conservative 200 nanoday period error estimate, that is still a factor of 2.90 times
smaller than 50 ms.
The period error estimates of \cite{Nemec+2011} are the average of period estimates 
obtained by the three
methods available with the software package Period04 
\citep{LenzBreger2004,LenzBreger2005}.  
The averaging of results from three different methods is not a proper error analysis ---
all three methods might be driven by the data to produce period estimates with more digits of precision
than are significant.  
The PEC algorithm predicts a period error of 2.6 microdays  (220 ms) for an uninterrupted 406-day 
\citep{KeplerDataReleaseNotesQ0toQ4,KeplerDataReleaseNotesQ5toQ8}
time series
and a timing uncertainty (half-bin width) of 0.0102 days. The lower limit for the PEC algorithm period error estimate 
is 531 nanodays ($\approx2.6\,{\rm{microdays}}/\sqrt{24}$) which 
is just below the 50 ms timing uncertainty for Kepler.

In a search for third star companions in 41 eclipsing binaries observed by the Kepler Mission,
\cite{Gies+2012} give the properties of these systems in their Table 1.
The periods and period error estimates were based on all the
long cadence Kepler light curve observations through Kepler quarter Q9. 
Only 2 of the 41 systems had period error estimates greater than
the 50 ms timing uncertainty for Kepler:
the best period error estimate is 2 nanodays and the worst is 20 microdays;
the median is 30 nanodays.  
A period estimate of 1.404678238 days with an uncertainty of 8 nanodays
was given for the primary star of the eclipsing binary KID 2305272.  
The PEC algorithm gives a period error estimate of 4.0 microdays for a uninterrupted 803-day 
\citep{,KeplerDataReleaseNotesQ9,KeplerDataReleaseNotesQ0toQ4,KeplerDataReleaseNotesQ5toQ8}
time series
and a timing uncertainty (half-bin width) of 0.0102 days. In order to achieve a precision 
of 8 nanodays with
the PEC algorithm, one needs a timing uncertainty of about 20 microdays (1.7 s) on every
Kepler observation. That timing accuracy is at
the very bottom of the range of internal timing measurement errors (1 to 318 s) reported by  \cite{Gies+2012}.
With 67.2 Kepler long cadence observations during a single
cycle (period) of KID 2305272, the lower limit for the PEC algorithm period error estimate 
is 0.5  microdays ($\approx4.0\,{\rm{microdays}}/\sqrt{67}$) which is below the 50 ms limit.
The timings used by \cite{Gies+2012} were based on model
fitting of parabolas to the lowest 20\% to the ``eclipse template data'' in order to find the actual phase of the
eclipse minimum.  It is not clear how the eclipse template data was mitigated to account for the long
exposures times of $\sim\,$1765.5 s.

The period errors estimated by the PEC algorithm are not the best possible period errors that can be
obtained for eclipsing binary systems.  If a lot more effort is expended, then better period error
estimates can usually be determined. 
The best period error estimates for the orbital periods of eclipsing binary systems are generally
determined using
least-squares model-fitting techniques on precision observations of photometry and radial velocities.
By comparing relative fluxes and radial velocities to physically detailed models of an eclipsing binary system,
all photometric and radial velocity information about the system can be optimized simultaneously.  For example,
Van Hamme \& Wilson's
(\citeyear{VanHammeWilson2007}) study
of the bright ($7.71\!\leq\!V\!\leq\!8.48$ mag; Kukarkin et al. \citeyear{Kukarkin+1971})
triple-system (Hilditch et al. \citeyear{Hilditch+1986}) Algol-type eclipsing binary DM Persei
yielded
a period error estimate of 180 nanodays by 
using intermittent observations over 
decades of ground-based photometry and radial velocity measurements:
$P_0 = 2.72774109(18)$ days.
\cite{Mikulasek+2012}
have presented a promising method for the period analysis of eclipsing binaries
which does not use $O-C$ diagrams.
This method was used by \cite{Zhu+2012} to determine the period of the 
relatively bright ($11.0\!\leq\!V\!\leq\!11.6$ mag; Kukarkin et al. \citeyear{Kukarkin+1971})
close eclipsing binary 
BS Vulpecula with a period error estimate of 20 nanodays:
$P_0 = 0.47597002(3)$ days.
Whether such precision is truly plausible
remains to be seen when this new method is applied in the future to other eclipsing binary systems.

\acknowledgements

This study was 
supported by NASA grant NNX10AC52G.
This paper includes data collected by the Kepler Mission. 
Kepler was competitively selected as the tenth Discovery
mission. Funding for this mission is provided by NASA's
Science Mission Directorate. 
This research has made use of the NASA Exoplanet Archive Periodogram Service, 
which is operated by the California Institute of Technology, 
under contract with the National Aeronautics and Space Administration under the 
Exoplanet Exploration Program.
Some of the data presented in this paper were obtained from the Mikulski Archive for 
Space Telescopes (MAST). STScI is operated by the Association of Universities for Research in 
Astronomy, Inc., under NASA contract NAS5-26555. Support for MAST for non-HST data is 
provided by the NASA Office of Space Science via grant NNX09AF08G and by other grants 
and contracts.
This research has made use of NASA's Astrophysics Data System Bibliographic Services.

{\em{Facility: \facility{Kepler}}}

\newpage



\newpage
\begin{deluxetable}{cll}
\tabletypesize{\scriptsize}
\tablecaption{Square of Period Error Estimates for 1 to 7 Cycles\label{tbl-one}}
\tablewidth{0pt}
\tablehead{
\colhead{M}
&\multicolumn{1}{l}{Expanded}
&\colhead{
Reduced}
}
\startdata
1
& 
$
\displaystyle
(\s{2}+\s{1})
$
& 
$
\displaystyle
2\sigma^2
$
\\[.8em]
\hline
\\[-1em]
2
& 
$
\displaystyle
\frac{1}{
2^2}
(\s{3}+\s{1})
$
& 
$
\displaystyle
\frac{\sigma^2}{2}
$
\\[1em]
\hline
\\[-1em]
3
& 
$
\displaystyle
\frac{1}{
3^2}
(\s{4}+\s{1})
$
& 
$
\displaystyle
\frac{2\sigma^2}{9}
$
\\[1em]
\hline
\\[-1em]
4
& 
\mbox{
$
\displaystyle
\frac{1}{2}
\bigg[
\,\frac{1}{
4^2}
(\s{5}+\s{1})
$
}
& 
$
\displaystyle
\frac{\sigma^2}{2}
\bigg[
\frac{1}{8}
$
\\[0.8em]
& 
\mbox{
$
\displaystyle
~~+
\frac{1}{2^2\times3^2}
\Big(
(\s{5}+\s{2})
+
(\s{4}+\s{1})
\Big)
\bigg]
$
}
&
$
\displaystyle
~~~~+
\frac{1}{9}
\bigg]
$
\\[1.1em]
\hline
\\[-1em]
5
& 
\mbox{
$
\displaystyle
\frac{1}{2}
\bigg[
\,\frac{1}{
5^2}
(\s{6}+\s{1})
$
}
& 
$
\displaystyle
\frac{\sigma^2}{2}
\bigg[
\frac{2}{25}
$
\\[0.8em]
& 
\mbox{
$
\displaystyle
~~+
\frac{1}{2^2\times4^2}
\Big(
(\s{6}+\s{2})
+
(\s{5}+\s{1})
\Big)
\bigg]
$
}
&
$
\displaystyle
~~~~+
\frac{1}{16}
\bigg]
$
\\[1.1em]
\hline
\\[-1em]
6
& 
\mbox{
$
\displaystyle
\frac{1}{3}
\bigg[
\,\frac{1}{
6^2}
(\s{7}+\s{1})
$
}
& 
$
\displaystyle
\frac{\sigma^2}{3}
\bigg[
\frac{1}{18}
$
\\[0.8em]
& 
\mbox{
$
\displaystyle
~~+
\frac{1}{2^2\times5^2}
\Big(
(\s{7}+\s{2})
+
(\s{6}+\s{1})
\Big)
$
}
&
$
\displaystyle
~~~~+
\frac{1}{25}
$
\\[1em]
& 
\mbox{
$
\displaystyle
~~+
\frac{1}{3^2\times4^2}
\Big(
(\s{7}+\s{3})
+
(\s{6}+\s{2})
+
(\s{5}+\s{1})
\Big)
\bigg]
$
}
&
$
\displaystyle
~~~~+
\frac{1}{24}
\bigg]
$
\\[1.1em]
\hline
\\[-1em]
7
& 
\mbox{
$
\displaystyle
\frac{1}{4}
\bigg[
\,\frac{1}{
7^2}
(\s{8}+\s{1})
$
}
& 
$
\displaystyle
\frac{\sigma^2}{4}
\bigg[
\frac{2}{49}
$
\\[0.8em]
& 
\mbox{
$
\displaystyle
~~+
\frac{1}{2^2\times6^2}
\Big(
(\s{8}+\s{2})
+
(\s{7}+\s{1})
\Big)
$
}
&
$
\displaystyle
~~~~+
\frac{1}{36}
$
\\[1em]
& 
\mbox{
$
\displaystyle
~~+
\frac{1}{3^2\times5^2}
\Big(
(\s{8}+\s{3})
+
(\s{7}+\s{2})
+
(\s{6}+\s{1})
\Big)
$
}
&
$
\displaystyle
~~~~+
\frac{2}{75}
$
\\[1em]
& 
\mbox{
$
\displaystyle
~~+
\frac{1}{4^2\times4^2}
\Big(
(\s{8}+\s{4})
+
(\s{7}+\s{3})
+
(\s{6}+\s{2})
+
(\s{5}+\s{1})
\Big)
\bigg]
$
}
&
$
\displaystyle
~~~~+
\frac{1}{32}
\bigg]
$
\enddata
\end{deluxetable}
\clearpage

\newpage
\begin{deluxetable}{crccrll}
\tablecaption{Period Errors of Variables in the Kepler Field\label{tbl-2}}
\tablewidth{0pt}
\tablehead{
\colhead{Kepler}
&\colhead{KID}
&\colhead{Type}
&\colhead{KEPMAG}
&\colhead{$P_{\rm{KEBC}}$}
&\colhead{$P$}
&\colhead{$\sigma_{\rm{PEC}}$}
\\
\colhead{Quarters}
&
&
&\colhead{(mag)}
&\colhead{(days)}
&\colhead{(days)}
&\colhead{(days)}
}
\startdata
Q0--Q2
& 11560447
& SD
& 10.834
& 0.527680
& 0.527678(39)
& 0.000014
\\
Q0--Q2
& 10858720
& SD
& 10.971
& 0.952386
& 0.952372(63)
& 0.000033
\\
Q0--Q2
& 9873869
& D
& 13.038
& 4.994774
& 4.99477(57)
& 0.00034
\\
Q0--Q2
& 3120320
& D
& 10.885
& 10.265600
& 10.2656(28)
& 0.00091
\\
Q0--Q2
& 9172506
& D
& 12.009
& 50.440245
& 50.442(19)
& 0.0072
\\[0.25em]
\hline
\\[-0.9em]
Q0--Q2
& 9641031
& D
& 9.177
& 2.178152
& 2.17816(13)
& 0.00011
\\
Q1--Q2
& 9540450
& D
& 14.146
& 2.154687
& 2.15472(26)
& 0.00011
\\[0.25em]
\hline
\\[-0.9em]
Q2
& 6936115
& RR Lyr
& 12.876
& \nodata
& 0.5273989(44)
& 0.000023
\\
Q2--Q8
& 6936115
& RR Lyr
& 12.876
& \nodata
& 0.5273989(14)
& 0.0000014
\enddata
\label{measurements}
\end{deluxetable}


\clearpage
\newpage

\clearpage
\newpage

\clearpage
\newpage

\clearpage
\newpage

\clearpage
\newpage

\end{document}